\documentclass[12pt]{article}
\usepackage[utf8]{inputenc}
\usepackage{geometry}
\usepackage{graphicx}
\usepackage{array}
\usepackage{subfig}
\usepackage{slashed}
\usepackage{amsmath}
\usepackage{amsfonts}
\usepackage{amssymb}
\usepackage[cm]{fullpage}
\usepackage{cite}
\usepackage{epstopdf}
\usepackage{comment}
\usepackage{hyperref}
\usepackage{cancel}
\usepackage{mathrsfs}

\usepackage{cleveref}
\crefname{section}{§}{§§}
\Crefname{section}{§}{§§}

\usepackage{booktabs} 

\numberwithin{equation}{section}

\def\p{\partial}

\def\0{{(0)}}
\def\1{{(1)}}
\def\2{{(2)}}

\def\<{\langle }
\def\>{\rangle }

\newcommand{\bea}{\begin{eqnarray}}

\newcommand{\eea}{\end{eqnarray}}

\newcommand{\be}{\begin{equation}}
\newcommand{\ee}{\end{equation}}
\newcommand{\ba}{\begin{align}}
\newcommand{\ea}{\end{align}}

   \makeatletter
  \let\over=\@@over \let\overwithdelims=\@@overwithdelims
  \let\atop=\@@atop \let\atopwithdelims=\@@atopwithdelims
  \let\above=\@@above \let\abovewithdelims=\@@abovewithdelims
\renewcommand\section{\@startsection {section}{1}{\z@}%
                                   {-3.5ex \@plus -1ex \@minus -.2ex}
                                   {2.3ex \@plus.2ex}%
                                   {\normalfont\large\bfseries}}

\renewcommand\subsection{\@startsection{subsection}{2}{\z@}%
                                     {-3.25ex\@plus -1ex \@minus -.2ex}%
                                     {1.5ex \@plus .2ex}%
                                     {\normalfont\bfseries}}

\newcommand{\beq}{\begin{equation}}
\newcommand{\eeq}{\end{equation}}
\newcommand{\beqa}{\begin{eqnarray}}
\newcommand{\eeqa}{\end{eqnarray}}
\newcommand{\beqar}{\begin{eqnarray*}}

\def\[{\[}
\def\]{\]}

\newcommand{\bd}[1]{\begin{fmffile}{#1}\begin{fmfgraph*}}
\newcommand{\ed}{\end{fmfgraph*}\end{fmffile}}

\begin{document}

\begin{titlepage}

\begin{flushright}
\end{flushright}

\unitlength = 1mm~\\
\vskip 1cm
\begin{center}

{\LARGE{\textsc{Gravitational Threshold Corrections  \\[0.3cm] in Non-Supersymmetric Heterotic Strings}}}

\vspace{1cm}
{\large Ioannis Florakis}

\vspace{1cm}

{\it     Laboratoire de Physique Th\'eorique et Hautes Energies,\\ Sorbonne Universit\'es, CNRS UMR 7589 - UPMC Paris 6, \\ 4 place Jussieu, 75005 Paris, France \\ 
		{\small {\tt florakis@lpthe.jussieu.fr}}

}

\vspace{0.8cm}

\begin{abstract}
We compute one-loop quantum corrections to gravitational couplings in the effective action of four-dimensional heterotic strings where supersymmetry is spontaneously broken by Scherk-Schwarz fluxes. We show that in both heterotic and type II theories of this class, no moduli dependent corrections to the Planck mass are generated.  We explicitly compute the one-loop corrections to the $\mathcal R^2$ coupling and find that, despite the absence of supersymmetry, its contributions may still be  organised into representations of subgroups of the modular group, and admit a universal form, determined uniquely by the multiplicities of the ground states of the theory. Moreover, similarly to the case of gauge couplings, also the gravitational sector may become strongly coupled in models which dynamically induce large volume for the extra dimensions.

\end{abstract}

\setcounter{footnote}{0}

\vspace{1.0cm}

\end{center}

\end{titlepage}

\pagestyle{empty}
\pagestyle{plain}

\def\vx{{\vec x}}
\def\p{\partial}
\def\po{$\cal P_O$}

\pagenumbering{arabic}

\bibliographystyle{utphys}

\section{Introduction}

Supersymmetry is undoubtedly one of the central ingredients of String Theory. Many of its spectacular results, such as duality connections, or various non-renormalisation theorems for string amplitudes,  rely heavily on its presence. On the other hand, our low energy world does not follow this simple paradigm and the eventual  spontaneous breaking of supersymmetry appears to be a necessary step for realistic string model building.

In the majority of cases in the string literature, one often works in a fully supersymmetric framework and postpones the breaking to a later stage of the analysis, at the level of the effective field theory. Nevertheless, quantitative comparisons to low energy data  necessitate the incorporation of quantum corrections to various couplings in the effective action, which generically receive sizeable contributions from the infinite tower of string states running in the loops. Including such corrections essentially calls for supersymmetry to be spontaneously broken from the very beginning, in the framework of a fully-fledged string theory. 

A special way to break supersymmetry spontaneously in String Theory that still admits an exact CFT description is the stringy uplift \cite{Rohm:1983aq,Kounnas:1988ye,Ferrara:1988jx,Kounnas:1989dk}  of the Scherk-Schwarz mechanism \cite{Scherk:1978ta,Scherk:1979zr}. In field theory, this amounts to a deformation of the theory in which a non-trivial monodromy is introduced on the various fields $\phi(x)$ as one encircles a compact dimension $x\sim x+2\pi R$. Namely, one imposes that fields in the theory are only periodic up to an isometry operation $Q$, namely $\phi(x+2\pi R)=e^{iQ}\,\phi(x)$. These modified boundary conditions have the effect of shifting the Kaluza-Klein spectra of charged states and, in particular, introduce a mass gap proportional to $Q/R$. If the original theory is supersymmetric, and one identifies $Q$ with the spacetime fermion number $F$, the deformed theory assigns different masses to states within the same supermultiplet, and supersymmetry is broken. From the point of view of the worldsheet theory, it manifests itself as a freely-acting orbifold correlating a translation along the compact direction $x$ of the internal manifold with a rotation in the internal super-coordinates that acts non-trivially on the R-symmetry lattice of the theory. 

A generic feature of this construction is that the supersymmetry breaking scale is tied to the size $R$ of the compact dimension along which one is shifting, $m_{3/2}\sim 1/R$. From the point of view of field theory, the Scherk-Schwarz mechanism is realised as a particular gauging of supergravity generating a tree-level scalar potential inducing masses to charged fields, while preserving chargeless fields, such as the Scherk-Schwarz radius $R$, massless (no scale moduli) \cite{Cremmer:1983bf}. The situation changes drastically if one considers quantum corrections to the scalar  potential. The latter are non-trivial functions of the no-scale moduli, and may lead to the stabilisation of some of them, or even introduce runaway behaviour for others \cite{Florakis:2016ani}. The fate of such no-scale moduli highly depends on the details of the particular string construction, which in turn determine the precise form of loop corrections to the effective potential. 

Although non-supersymmetric string constructions have been considered extensively in the literature\cite{AlvarezGaume:1986jb,Dixon:1986iz,Ginsparg:1986wr,Ferrara:1987es,Nair:1986zn,Itoyama:1986ei,Taylor:1987uv,Toon:1990ij,Dienes:1995bx,Sasada:1995wq,Blum:1997cs,Blum:1997gw,Harvey:1998rc,Ghilencea:2001bv,Font:2002pq,Faraggi:2007tj,Florakis:2009sm,Faraggi:2009xy,Florakis:2010ty,Florakis:2010is,Angelantonj:2006ut,GatoRivera:2007yi,GatoRivera:2008zn}, \cite{Blaszczyk:2014qoa,Abel:2015oxa,Lukas:2015kca,Ashfaque:2015vta,Blaszczyk:2015zta,Nibbelink:2015vha}, the analysis of loop corrections to couplings has only began fairly recently \cite{Angelantonj:2014dia,Florakis:2015txa,Faraggi:2014eoa,Kounnas:2016gmz,Angelantonj:2015nfa,Angelantonj:2016ibb}. In \cite{Angelantonj:2014dia,Angelantonj:2015nfa}, the one-loop running of gauge couplings was explicitly computed in a generic class of heterotic orbifold compactifications in which supersymmetry was spontaneously broken \`a la Scherk-Schwarz.  The main result of this analysis is that the moduli dependent contribution to threshold differences is governed by an unexpected universality structure
\begin{align}
	\begin{split}
	\Delta  = &\sum_{i=1,2,3} \left\{ \alpha_i \,\log\left[ T_2^{(i)} U_2^{(i)}\,\left| \eta(T^{(i)})\,\eta(U^{(i)})\right|^4 \right]  \right. \\
	&\left. +\beta_i \,\log\left[ T_2^{(i)} U_2^{(i)}\,\left| \vartheta_4(T^{(i)})\,\vartheta_2(U^{(i)})\right|^4 \right] 
					 + \gamma_i \,\log\left|j_2(T^{(i)}/2)-j_2(U^{(i)})\right|^4 \right\} \,,
	\end{split}
	\label{gaugeuniversality}
\end{align}
where $\Delta$ is the difference of one-loop threshold corrections for two gauge group factors, while $T^{(i)}, U^{(i)}$ are the K\"ahler and complex structure moduli of the $i$-th 2-torus inside the $T^6$ internal space, and $\eta(\tau),\vartheta(\tau)$ denote the Dedekind and Jacobi theta functions, respectively. Moreover, $j_2(\tau)=(\eta(\tau)/\eta(2\tau))^{24}+24$ is the Hauptmodul for the $\Gamma_0(2)$ congruence subgroup of ${\rm SL}(2;\mathbb Z)$ and is the analogue of the Klein invariant $j$-function of the full modular group. In this expression, the details on the particular model are entirely absorbed into the constant prefactors $\alpha, \beta,\gamma$ which are in turn expressible in terms of differences of beta function coefficients, and are easily computable from the massless spectrum of the theory. 

As soon as supersymmetry is broken spontaneously in string theory one is immediately asked to face two fundamental problems. The first concerns the possibility of encountering tachyonic modes in the string spectrum and is inherently related to the exponential growth of the degeneracy of states of string theory \cite{Atick:1988si,Axenides:1987vi}. It manifests itself as the Hagedorn instability of string theory in a thermal setup\cite{Antoniadis:1991kh,Antoniadis:1999gz,Kutasov:1990sv,Dienes:1994np,Angelantonj:2008fz,Angelantonj:2010ic}. The second, is related to the presence of a one-loop tadpole \cite{Fischler:1986ci,Fischler:1986tb} and is basically a back-reaction problem. Interesting exceptions do exist for special constructions for which supersymmetry is  broken but the one-loop vacuum energy either vanishes identically \cite{Kachru:1998hd,Kachru:1998pg,Harvey:1998rc,Shiu:1998he,Angelantonj:1999gm} or is exponentially suppressed \cite{Abel:2015oxa,Kounnas:2016gmz,Florakis:2016ani}.

In a very recent work \cite{Florakis:2016ani},  explicit examples of chiral heterotic models were presented in which the quantum corrections to the scalar potential actually induce a spontaneous decompactification of the Scherk-Schwarz radius, leading to its dynamical roll into the large volume regime $R\gg \ell_s$, where the supersymmetry breaking scale is suppressed, while maintaining an exponentially small but positive value for the cosmological constant. An additional advantage is that these theories are dynamically secured against the excitation of tachyonic modes. Given the possibility of explicitly realising such scenarios involving large extra dimensions \cite{Antoniadis:1990ew} within a stringy framework, it is important and necessary to extend the study of quantum corrections to other couplings in the string effective action and especially to extract their dependence on the Scherk-Schwarz moduli. 

One of the issues one faces in heterotic string models of this type is the fact that, as one may verify by inspecting the first term on the r.h.s. of eq. \eqref{gaugeuniversality},  the threshold correction to gauge couplings with negative $\mathcal N=2$ beta function coefficient is quickly driven into the strong coupling regime, as the volume of the extra dimensions becomes much larger than the string scale. Namely, as the volume $T_2\equiv {\rm Im}(T)$ of the Scherk-Schwarz 2-torus becomes large, the physics becomes effectively six-dimensional and the coupling is dominated by a linear growth in the volume $T_2$, which replaces the logarithmic four-dimensional one. This is known as the decompactification problem of gauge thresholds \cite{Kiritsis:1996xd,Faraggi:2014eoa}.

In this spirit, the purpose of this paper is to examine the structure of one-loop threshold corrections to gravitational $\mathcal R^2$ couplings in heterotic theories with spontaneously broken supersymmetry\footnote{In supersymmetric setups quantum corrections to $\mathcal R^2$ couplings have been extensively studied in the past, c.f. \cite{Antoniadis:1992sa,Antoniadis:1992rq,Kiritsis:1994ta,Vafa:1995fj,Harvey:1996ir,Forger:1996vj,Gregori:1997hi}. See also the more recent generalisation \cite{Angelantonj:2016gkz}.}. To this end, we employ the background field method of \cite{Kiritsis:1994ta} which involves consistently deforming the flat non-supersymmetric theory by switching on an exact (1,1) integrable perturbation associated to a non-trivial but constant curvature $\mathcal R$  and obtaining the exact deformed partition function $Z(\mathcal R)$. The latter may be seen as a generating function that, when expanded around the flat background at a given order, allows one to extract the one-loop renormalisation of the gravitational couplings of interest. 

We show that the Planck mass receives no moduli-dependent corrections in either heterotic or type II string theories with spontaneously broken supersymmetry. What is more, we find that moduli-independent corrections to the $\mathcal R$ term may arise and are precisely identical to the supersymmetric case. We further find that one-loop corrections to the $\mathcal R^2$ term are non-trivial  and may even lead the theory into a strong coupling regime in models that dynamically favour a large volume of their internal space. In this way, the decompactification problem of gauge couplings and its gravitational counterpart furnish important constraints on model building.

In the absence of supersymmetry, gravitational $\mathcal R^2$ thresholds no longer display the simple holomorphy properties of the supersymmetric BPS-protected counterparts, but are nevertheless explicitly calculable. We show that their contributions can be cast as stringy Schwinger-like integrals organised into representations of the Hecke congruence subgroup $\Gamma_0(2)$  of ${\rm SL}(2;\mathbb Z)$ and are expressible in a universal form. The only model dependence is absorbed into a set of beta function coefficients that can be uniquely determined from the knowledge of multiplicities of the ground states of the theory.

The paper is structured as follows. In section \ref{SecBFM}, we review the background field method allowing us to extract unambiguously the general form of one-loop $\mathcal R$ and $\mathcal R^2$ corrections. In section \ref{PlanckMass} we employ this machinery in order to discuss the absence of moduli-dependent renormalisation of the Planck mass for both heterotic and type II theories with spontaneously broken supersymmetry. In section \ref{SecR2thresh}, we assemble the necessary ingredients for the evaluation of the gravitational $\mathcal R^2$ thresholds by determining the necessary correlators as trace insertions in the deformed partition function of the theory and then move on to compute them in a specific prototype model in section \ref{prototypeSection}. Finally, in section \ref{SecLargeVol} we discuss the universal properties of gravitational thresholds and extract their large volume behaviour, in connection with the models recently constructed in \cite{Florakis:2016ani}.

\section{Background field method}\label{SecBFM}

A very efficient method for extracting the one-loop corrections to the couplings of interest is the background field method. The idea is to deform the worldsheet sigma model by an operator of the form
\begin{equation}
	\delta S_{\rm het}=\int d^2 z  \,\mathcal R ( X^{[1} \partial X^{2]} + \psi^1 \psi^2)\,X^{[1}\bar\partial X^{2]}\,,
	\label{origDeform}
\end{equation}
corresponding to a small, constant curvature $\mathcal R$ and evaluate the partition function of the deformed theory $Z(\mathcal R)$. Here, $\psi^1, \psi^2$ are worldsheet fermions in the corresponding spacetime directions and $X^1, X^2$ are their bosonic super-partners. Correlation functions of interest are then obtained as appropriate derivatives of the generating function $Z(\mathcal R)$, around the trivial background $\mathcal R=0$. Unfortunately, the above deformation is not marginal in flat space, due to the fact that the rotation operators $X^{[\mu}\partial X^{\nu]}$ and $X^{[\mu}\bar\partial X^{\nu]}$ are not well-defined conformal fields. In other words, the above insertion induces a non-trivial back-reaction.

Following \cite{Kiritsis:1994ta}, this situation can be remedied by replacing the flat four dimensional spacetime by the $W_k^{(4)}$ space. This is an $\hat c=4$ superconformal system involving a 
 supersymmetric ${\rm SU}(2)_k$ WZW model as well as a non-compact dimension with background charge $Q=\sqrt{\frac{2}{k+2}}$, such that in the limit of small curvature (large level $k$), the $W_k^{(4)}$ space can be approximated as the space $\mathbb{R} \times S^3$. Assuming $k$ is even, the partition function in the $W_k^{(4)}$ theory $Z_W$ is readily obtained from the original one in flat space $Z_0$ that corresponds to our heterotic theory of interest, as
\begin{equation}
	Z_W = \Gamma_{{\rm SU}(2)_k}\, Z_0 \,.
\end{equation}
Here, $\Gamma_{{\rm SU}(2)_k}$ is the partition function of the worldsheet bosons in the $W_k^{(4)}$ space divided by their flat space contribution
\begin{equation}
	\Gamma_{{\rm SU}(2)_k} = (\sqrt{\tau_2}\,\eta\bar\eta)^3 \,\frac{1}{2}\sum_{\alpha,\beta=0,1} e^{-i\pi k\alpha\beta/2}\sum_{\ell=0}^k e^{i\pi \beta\ell}\, \chi_{\ell}(\tau)\,\bar\chi_{\ell+\alpha(k-2\ell)}(\bar\tau) \,,
\end{equation}
and $\chi_\ell(\tau)$ are the affine characters of ${\rm SU}(2)_k$ with spin $j=\ell/2$. They are holomorphic in the complex structure $\tau=\tau_1+i\tau_2$ on the worldsheet torus and can be expressed in terms of level-$k$ theta functions according to the parafermionic decomposition (see for instance \cite{Gepner:1986hr}). The leading behaviour at large volume $k\gg 1$ is now proportional to $e^{-\frac{\pi\tau_2}{k+2}}$ and correctly indicates the presence of a mass gap due to the effective replacement of the flat $\mathbb R^3$ space by a 3-sphere.

The advantage of working with the $W_k^{(4)}$ space is that we can now consistently deform the worldsheet sigma model of our heterotic theory by the exact (1,1) operator
\begin{equation}
	\delta S_{\rm het} = \int d^2 z\, \mathcal R \left( J^3+\psi^1\psi^2 \right)\, \tilde J^3 \,,
	\label{W4deform}
\end{equation}
which, in the flat limit corresponds to a perturbation with constant Riemann tensor, and with corresponding $B$-field and dilaton profiles, all proportional to the perturbation constant $\mathcal R$. Here, $J^3$ is the diagonal Kac-Moody current $J^3=k\,{\rm Tr}(\sigma^3 g^{-1}\partial g)$ of ${\rm SU}(2)_k$, whose group elements are parametrised as $g=e^{i\sigma\cdot X/2}$, and similarly for the right movers. Expanding the group element $g$ to quadratic order in the coordinates, one obtains $J^3 \sim k(\partial X^3 + X_{[1}\partial X_{2]}+\ldots)$, matching the desired deformation \eqref{origDeform}.

The deformation \eqref{W4deform} is exactly integrable and can be recognised as a boost of the fermionic and SU(2) charge lattices \cite{Kiritsis:1994ta}
\begin{equation}
	\delta L_0 = \mathcal R\frac{(Q+J^3_0)\bar J^3_0}{\sqrt{k(k+2)}}+ \frac{\sqrt{1+\mathcal R^2}-1}{2} \,\frac{(Q+J^3_0)^2}{k+2} + \frac{\mathcal R^2}{1+\sqrt{1+\mathcal R^2}}\,\frac{(\bar J_0^3)^2}{k}\,,
\end{equation}
and similarly for the right-moving part. Here $\delta L_0$ is the deformation of the Virasoro zero mode, $J_0^3$ is the zero mode of the Kac-Moody current $J^3(z)$ and $Q=\oint J_{12}$ is the U(1) helicity charge ascribed to the worldsheet fermion current, $J_{12} = \psi^1 \psi^2$. By the very definition of the boost, $\delta L_0 - \delta \bar L_0=0$ and, therefore, derivatives of the deformed partition function with respect to $\mathcal R$ evaluated at $\mathcal R=0$ yield 
\begin{equation}
	\begin{split}
	&\langle \mathcal R\rangle  = -4\pi \tau_2 \left\langle (Q+J_0^3)\,\bar J_0^3  \right\rangle \,,\\
	&\langle \mathcal R^2\rangle  = \left\langle 8\pi^2 \tau_2^2\left( (Q+J_0^3)^2 - \frac{k+2}{8\pi\tau_2}\right) \left( (\bar J_0^3)^2 - \frac{k}{8\pi\tau_2}\right) -\frac{k(k+2)}{8}\right\rangle\,.
	\end{split}
	\label{insertions}
\end{equation}
The above generic expressions are valid regardless of whether supersymmetry is spontaneously broken or not. They identify the relevant zero-mode insertions which should be then weighted by the remaining contributions in the partition function $Z_W$ of the curved $W_k^{(4)}$ space and eventually integrated over the moduli space of the complex structure $\tau$ of the worldsheet torus in order to yield physical couplings. Of course, since we are interested in the renormalisation of gravitational couplings in flat space, we will eventually take the flat space limit $k\to \infty$.


\section{Non-renormalisation of the Planck mass}\label{PlanckMass}


Before embarking on the calculation of $\mathcal R^2$ threshold corrections, it is first instructive to comment on the fact that the Planck mass does not receive moduli-dependent renormalisation at one-loop in either heterotic or type II string theory, despite the possible spontaneous breaking of supersymmetry by the Scherk-Schwarz mechanism.

\subsection{Heterotic theories}

We begin with the heterotic case. From \eqref{insertions}, the relevant terms are $\langle Q \bar J_0^3\rangle$ and $\langle J_0^3\bar J_0^3\rangle$. Both involve SU(2) charge insertions and can be evaluated using the $z$-deformed characters
\begin{equation}
	\chi_\ell(z,\tau) = {\rm Tr}_{(\ell)}\left[\, q^{L_0-\frac{c}{24}}\,e^{2\pi i \langle z, J^3\rangle } \,\right] = 2\sum_{\gamma\in\Lambda_\ell} q^{(k+2)\langle\gamma,\gamma\rangle}\,\frac{\sin(2\pi(k+2) \langle z,\gamma\rangle)}{\vartheta_1(z,\tau)} \,,
	\label{characters}
\end{equation}
with $q=e^{2\pi i \tau}$ and $\gamma$ summed over the affine lattice. For the term $\langle Q \bar J_0^3\rangle$, taking a single $z$-derivative in $\bar\chi$ and evaluating the resulting expression around $z=0$, produces a vanishing result proportional to $\bar\vartheta_1''(0)/[\bar\vartheta_1'(0)]^2=0$. Performing a similar analysis for the term $\langle J_0^3\bar J_0^3\rangle$  again yields a vanishing result for the same reason, expressing the charge conservation of ${\rm SU}(2)_{L,R}$. The presence or not of spacetime supersymmetry has played no role in this argument, which is generic in heterotic theories at one and higher genera \cite{Kiritsis:1994ta}.

\subsection{Type II theories}

In the type II case, using again the background field method, one instead identifies the relevant insertion to be $-2\pi\tau_2\langle (Q+J_0^3)(\bar Q+\bar J_0^3)\rangle$. Charge conservation again yields vanishing contributions for all but the term $\langle Q\bar Q\rangle$. The latter insertion corresponds to taking $z,\bar z$-derivatives with respect to the corresponding left- and right- moving theta functions associated to the worldsheet fermions in the spacetime directions, namely
\begin{equation}
	{\rm Tr}_{\mathcal H_a}\left[\,Q\, e^{-i\pi b Q}\,q^{L_0} \,\right]= \left.\frac{1}{2\pi i}\frac{\partial_z\vartheta[^a_b](z,\tau)}{\eta(\tau)} \right|_{z=0}\,,
	\label{zderiv}
\end{equation}
and similarly for the right movers. Consider now a generic consistent orbifold that preserves at least one spacetime supersymmetry from the left-movers and at least one from the right. It will have some crystallographic action on some or all of the complexified internal $T^6$ super-coordinates as
\begin{equation}
	Z^i \to e^{2\pi i v_i} \,Z^i \quad,\quad \Psi^i \to e^{2\pi i v_i} \,\Psi^i \,,
\end{equation}
with $i=1,2,3$. Let us also assume that we further deform the theory \`a la Scherk-Schwarz to completely break all remaining supersymmetries, by coupling the left- and right- moving R-symmetry charges to a shift in some internal lattice direction. The twisted partition function contribution to the worldsheet fermions is of the general form
\begin{equation}
	Z[^{H,h_i}_{G,g_i}]=\frac{1}{2}\sum_{a,b=0,1} (-1)^{a(1+H)+b(1+G)+HG}\, \,\vartheta[^a_b](z)\,\vartheta[^{a+h_1}_{b+g_1}](0)\,\vartheta[^{a+h_2}_{b+g_2}](0)\,\vartheta[^{a-h_1-h_2}_{b-g_1-g_2}](0)\,,
\end{equation}
and we have introduced the deformation $z$ in the first theta function corresponding to the fermions in the spacetime directions. Here, $H,G=0,1$ label the boundary conditions from the Scherk-Schwarz $\mathbb Z_2$ twist and $h_i,g_i$ similarly account for the fermion boundary conditions in the generic sector of the supersymmetry-preserving orbifold.  A similar expression holds for the right-movers and the supersymmetry preserving orbifold could even be asymmetric.
Performing the sum over the spin structures $(a,b)$ using the Riemann identity, we find
\begin{equation}
	Z[^{H,h_i}_{G,g_i}]= (-1)^{HG}\,\vartheta[^{1+H}_{1+G}](z/2)\,\vartheta[^{1+H+h_1}_{1+G+g_1}](z/2)\,\vartheta[^{1+H+h_2}_{1+G+g_2}](z/2)\,\vartheta[^{1+H-h_1-h_2}_{1+G-g_1-g_2}](z/2) \,.
	\label{RiemanIdent}
\end{equation}
If $H=G=0$, which corresponds to the case of unbroken supersymmetry, the $z$ derivative \eqref{zderiv} must necessarily act on the first theta function to soak up the zero modes and yields a non-vanishing contribution to the renormalisation of the Planck mass, provided all three remaining theta functions are twisted. This can occur only from the moduli-independent $\mathcal N=(1,1)$ sectors and the non-trivial one-loop contribution to the Planck mass was computed in \cite{Kiritsis:1994ta} and later generalised in \cite{Kohlprath:2002fe}.

Let us now consider the case with Scherk-Schwarz breaking $(H,G)\neq (0,0)$ for which the possibility of one-loop renormalisation to the Planck mass has not been studied previously in a type II setup. The $z$-derivative can potentially act only on the last three theta functions in \eqref{RiemanIdent}, since $\vartheta_{j}'(0)=0$ for $j=2,3,4$. Assume that the Scherk-Schwarz acts as a shift along the first $T^2$ torus, which is twisted by $(h_1,g_1)$. Moduli dependent contributions could only arise from $\mathcal N=(2,2)$ sectors, in which the twist along the direction of the first $T^2$ necessarily vanishes $h_1=g_1=0$. However, using the identity 
\begin{equation}
	\vartheta[^{-a}_{-b}](z,\tau) = \vartheta[^{a}_{b}](-z,\tau)\,,
\end{equation}
together with the periodicity formulas for theta functions, we again obtain
\begin{equation}
	 \vartheta[^{1+H}_{1+G}]^2(0)\ \partial_z\Bigr[\vartheta[^{1+H+h_2}_{1+G+g_2}](z/2)\,\vartheta[^{1+H-h_2}_{1+G-g_2}](z/2) \Bigr]_{z=0} =0\,,
\end{equation}
and there can be no moduli-dependent contributions to the Planck mass renormalisation even in the case of spontaneously broken supersymmetry.

 Consider now the moduli independent contributions arising from $\mathcal N=(1,1)$ sectors. In particular, now $(h_1,g_1)$, $(h_2,g_2)$ and $(h_1+h_2,g_1+g_2)$ are different than $(0,0)$ modulo 2.
Since there is also a non-trivial Scherk-Schwarz shift by $(H,G)\neq (0,0)$ in the first $T^2$, the simultaneously twisted and shifted Scherk-Schwarz lattice produces a non-vanishing contribution only if $(h_1,g_1)=(H,G)\,{\rm mod}\,2$. Returning to \eqref{RiemanIdent}, we find up to overall phases
\begin{equation}
 \sim \vartheta[^{1+H}_{1+G}](0)\,\vartheta'[^{1}_{1}](0)\,\vartheta[^{1+H+h_2}_{1+G+g_2}](0)\,\vartheta[^{1-h_2}_{1-g_2}](0) \,.
\end{equation}
The $z$-derivative must necessarily act on the second theta function and soaks up the zero modes, as required in order to yield a non-vanishing contribution. Naturally, since all three $T^2$ lattices are twisted, there is no moduli dependence from $\mathcal N=(1,1)$ sectors but, nevertheless, the renormalisation of the Planck mass is a non-vanishing constant, proportional to the volume of the fundamental domain
\begin{equation}
	\langle \mathcal R\rangle = c\int_{\mathcal F} \frac{d^2\tau}{\tau_2^2}\,\Gamma_{{\rm SU}(2)_k}= \frac{\pi c}{3}\left(1+\frac{2}{k+2}\right)\,,
\end{equation}
 similarly to the case considered in \cite{Kiritsis:1994ta}. Here, $c$ is the normalised  multiplicity of $\mathcal N=(1,1)$ sectors. In the flat space limit $k\to \infty$ it becomes simply $\langle \mathcal R\rangle = c\pi/3$. 
 
It is extremely interesting that the one-loop renormalisation of the Einstein-Hilbert term in type II theories is moduli independent, even in the case of spontaneous supersymmetry breaking, and has the precise same form as in supersymmetric theories. This result holds whenever there are $\mathcal N=(1,1)$ subsectors in the theory. In particular, it holds also for constructions with spontaneously broken $\mathcal N=1$ supersymmetry.


\section{$\mathcal R^2$ Threshold corrections} \label{SecR2thresh}

We now move on to the study of $\mathcal R^2$ thresholds which are generically non-vanishing. The starting point is to identify the insertions in \eqref{insertions}. Using the same argument of SU(2) charge conservation as in section \ref{PlanckMass}, linear terms in the $J_0^3$ current drop out, and we can consider instead
\begin{equation}
	\langle \mathcal R^2\rangle  = \left\langle 8\pi^2 \tau_2^2\left( Q^2+J^2 - \frac{k+2}{8\pi\tau_2}\right) \left( \bar J^2 - \frac{k}{8\pi\tau_2}\right)\right\rangle -\frac{k(k+2)}{8}\langle 1\rangle\,,
\end{equation}
where we shall henceforth drop the indices in the Kac-Moody current zero modes and simply denote $J_0^3, \bar J_0^3$ by $J$ and $\bar J$, respectively.
The $Q^2$ insertion is straightforward to compute using \eqref{zderiv} and the heat equation satisfied by theta functions, so that we may write
\begin{equation}
	\langle Q^2\rangle= \frac{1}{i\pi \eta(\tau)}\,\partial_\tau\vartheta[^a_b](0,\tau) \,.
\end{equation}
Let us now consider the $\langle J^2\rangle$ insertion. Taking two $z$-derivatives on \eqref{characters} at $z=0$ and turning the cubic charge insertion into a derivative with respect to $\tau$, allows one to express
\begin{equation}
	\langle J^2\rangle=\left(-\frac{1}{12}E_2(\tau)+\frac{k+2}{6\pi i}\,\partial_\tau + \frac{k+2}{8\pi\tau_2}\right)\Gamma_{{\rm SU}(2)_k}(\tau,\bar\tau)\,,
\end{equation}
where $E_2(\tau)$ is the holomorphic but almost modular Eisenstein series of weight 2.
We may therefore write
\begin{equation}
	\begin{split}
	\langle \mathcal R^2\rangle =& -8\tau_2^2 \left\langle \partial_\tau\log\left(\frac{\vartheta[^a_b]}{\eta}\right)\left(-\frac{i\pi}{12}\hat{\bar E}_2-\frac{k+2}{6}\partial_{\bar\tau}\right) \right. \\
	&\left. +\frac{k+2}{6}\,\left( -\frac{i\pi}{12}\hat{\bar E}_2-\frac{k+2}{6}\partial_{\bar\tau}\right)\partial_\tau\right\rangle\,\Gamma_{{\rm SU}(2)_k} -\frac{k(k+2)}{8}\,\Gamma_{{\rm SU}(2)_k}\,\langle 1\rangle\,,
	\end{split}
	\label{master}
\end{equation}
where $\hat E_2=E_2-3/\pi\tau_2$ is the almost holomorphic but modular Eisenstein series of weight 2, and with the understanding that the quantities inside the brackets $\langle\ldots\rangle$ are to be weighted using the original partition function of the theory and integrated over the moduli space of the worldsheet torus. In particular, the last term is proportional to the vacuum energy $\langle 1\rangle$ of the theory, which is non-vanishing in the case of spontaneously broken supersymmetry. However, this term is related to the choice of renormalisation scheme and, in particular, depends on the IR mass gap parameter $\mu=1/\sqrt{k+2}$, \cite{Kiritsis:1994ta}. It will be dropped in the following considerations\footnote{See \cite{Angelantonj:2011br,Angelantonj:2012gw} for the relation of this scheme to the one where the fundamental domain $\mathcal F$ is truncated.} and will be cancelled out in evaluation of physical observables provided the exact same regularisation is properly carried out in a low energy field theory analysis (see also \cite{Kiritsis:1994ta}).

These expressions are valid for the curved theory on $W_k^{(4)}$. To obtain the result of interest in flat space, we take the decompactification limit $k\to\infty$, and suitably divide by the SU(2) volume factor $V_k=(k+2)^{3/2}/8\pi$. In order to correctly extract this limit, we first perform the sum over $\ell$ in $\Gamma_{{\rm SU}(2)_k}$ and rewrite the level-$k$ theta functions as a bosonic charge lattice. The Lagrangian representation is the most suitable one for extracting the limit and reads
\begin{equation}
	\Gamma_{\rm {SU}(2)_k}(\tau,\bar\tau) = V_k \sum_{m,n\in\mathbb Z} (-1)^{m+n+mn}\left(1-\frac{\pi(k+2)}{2\tau_2}|m-\tau n|^2\right)e^{-\frac{\pi(k+2)}{4\tau_2}|m-\tau n|^2} \,.
\end{equation}
From this expression, one may readily observe that in the $k\to\infty$ limit, the dependence on $\tau,\bar\tau$ drops out and one is left with the SU(2) volume $V_k$ arising from the term $m=n=0$. This implies that terms involving $\tau$ or $\bar\tau$ derivatives acting on $\Gamma_{{\rm SU}(2)_k}$ in \eqref{master} may be dropped since they will be proportional to $m,n$ and, therefore, vanish in the flat space limit. The resulting expression for the $\mathcal R^2$ thresholds in flat space then reads
\begin{equation}
	\Delta_{\rm grav}= \frac{i\pi}{12}\,\int_{\mathcal F} d\mu\ \frac{\tau_2\,\hat{\bar E}_2}{\eta^2\bar\eta^2} \ \sum_{a,b} \left\langle \partial_\tau\log\left(\frac{\vartheta[^a_b]}{\eta}\right)\right\rangle \,,
	\label{gravthresh}
\end{equation}
where we have reinstated the integral of the Teichm\"uller parameter $\tau$ over the fundamental domain $\mathcal F=\mathbb H^+/{\rm SL}(2;\mathbb Z)$. The latter is obtained as the quotient of the Teichm\"uller space (upper half-plane $\mathbb H^+$) by the mapping class (modular) group. It is important to note that this general expression is identical to one we would have obtained in the supersymmetric case. This is a highly non-trivial statement which is possible to prove unambiguously using the background field method. A priori, there was  no reason to expect that the correlator insertions would have had the exact same form \eqref{gravthresh} in both the non-supersymmetric as well as in the supersymmetric case. After all, in curved space $W_k^{(4)}$ there are additional non-vanishing backreaction contributions in the absence of supersymmetry! It is only in the flat space limit that the additional terms are washed out and one recovers in both cases the same simple expression \eqref{gravthresh}.

Of course, the fact that the correlator insertion is the same in flat space does not mean that the actual thresholds will be identical in both supersymmetric and non-supersymmetric cases. The actual evaluation of the correlators depend on the choice of background. In the absence of supersymmetry, the evaluation of the weighted correlator $\langle\ldots\rangle$ will no longer possess simple quasi-holomorphy (BPS) properties and will typically be a manifestly non-holomorphic function of $\tau, \bar\tau$. In what follows, we shall evaluate $\Delta_{\rm grav}$ for a simple heterotic theory with $\mathcal N=2\to 0$ spontaneous breaking.


\section{An anatomy of $\mathcal R^2$ corrections in a prototype model}\label{prototypeSection}

It will be sufficient for our purposes to perform explicit computations in the model of ref. \cite{Angelantonj:2014dia} which is characterised by spontaneously broken $\mathcal N=2\to 0$ supersymmetry. It can be obtained as the Scherk-Schwarz reduction of the six-dimensional K3 compactification of the ${\rm E}_8\times{\rm E}_8$ heterotic string and
 has the advantage of being free of tachyonic modes at every point in the K\"ahler and complex structure $(T,U)$ moduli space associated to the 2-torus of the Scherk-Schwarz reduction, provided one does not turn on additional Wilson line deformations. 

This specific prototype model was chosen precisely due to its simplicity, so as to best display the salient features of the structure of $\mathcal R^2$ corrections, but it will also serve as a basis for generalisation to more realistic models. It was also in this model that the one-loop corrections to gauge thresholds were first computed in non-supersymmetric heterotic strings.

The model can be constructed as a $T^6/\mathbb Z_2\times \mathbb Z_2'$ orbifold, where the first $\mathbb Z_2$ factor has the standard crystallographic action $(Z^1,Z^2)\to (-Z^1,-Z^2)$ on the complexified (super-) coordinates of a $T^2\times T^2$ subspace, whereas the second $\mathbb Z_2'$ factor is freely-acting and involves a momentum shift along the third  $T^2$ space, correlated with a parity insertion involving the spacetime fermion number $F$ and the ``fermion numbers" $F_1, F_2$ associated with the spinorial representations of the two original ${\rm E}_8$ factors of the theory. Explicitly, it reads
\begin{equation}
	v'\in\mathbb Z_2' \ : \ \quad v'=(-1)^{F+F_1+F_2}\,\delta\,,
	\label{ScherkSchwarzAction}
\end{equation}
where $\delta$ is the order two shift along the direction of the remaining 2-torus, rendering the action of $\mathbb Z_2'$ free and the breaking of supersymmetry spontaneous.
The modular covariant one-loop partition function can be expressed in the following form
\begin{equation}
	\begin{split}
	Z_0=&\frac{1}{\eta^{12}\bar\eta^{24}}\,\frac{1}{2}\sum_{H,G=0,1}\frac{1}{2}\sum_{h,g=0,1}\frac{1}{2}\sum_{a,b=0,1}(-1)^{a+b}\,\vartheta[^a_b]^2\,\vartheta[^{a+h}_{b+g}]\,\vartheta[^{a-h}_{b-g}]\\
		&\times\frac{1}{2}\sum_{k,\ell=0,1} \bar\vartheta[^k_\ell]^6\,\bar\vartheta[^{k+h}_{\ell+g}]\,\bar\vartheta[^{k-h}_{\ell-g}]\, \frac{1}{2}\sum_{\rho,\sigma=0,1}\bar\vartheta[^\rho_\sigma]^8 \\
		& \times \Gamma_{4,4}[^h_g]\,\Gamma_{2,2}[^H_G]\, (-1)^{H(b+\ell+\sigma)+G(a+k+\rho)+HG}\,.
	\end{split}
	\label{prototype}
\end{equation}
Here, $H,h$ label the twisted sectors of the two $\mathbb Z_2$ orbifold factors, while summation over $G,g$ imposes the corresponding invariant projections. As before, the sum over the $(a,b)$ spin structures labels the spacetime bosons and fermions of the theory and imposes the GSO projection of the heterotic superstring. The purely right-moving contributions in the second line reconstruct the partition function of the ${\rm E}_8\times {\rm E}_8$ lattice, twisted by the $\mathbb Z_2$ orbifold with standard embedding. Before the Scherk-Schwarz breaking, it is identified with the chiral ${\rm E}_7\times {\rm SU}(2)\times{\rm E}_8$ lattice blocks associated to the level one Kac-Moody algebra and expressible entirely in terms of ordinary theta characters. The third line contains the twisted (4,4) lattice where the crystallographic K3-like action $(h,g)$ takes place and is given by
\begin{equation}
	\Gamma_{4,4}[^h_g] = \left\{ 
				\begin{array}{l l}
				\Gamma_{4,4}(G,B) & ,\ \ (h,g)=(0,0)\\
				 &\\
				\dfrac{16\eta^6\bar\eta^6}{\left|\vartheta[^{1+h}_{1+g}]\vartheta[^{1-h}_{1-g}]\right|^2} & ,\ \ (h,g)\neq(0,0)
				\end{array}
	\right. \,,
	\label{K3lattice}
\end{equation}
where $\Gamma_{4,4}(G,B)$ is the Narain lattice of $T^4$ 
\begin{equation}
	\Gamma_{4,4}(G,B) = \sum_{\vec m,\vec n} q^{\frac{1}{4}\vec P_L^2}\,\bar{q}^{\frac{1}{4}\vec P_R^2}\,,
	\label{44lattice}
\end{equation}
and depends on the hypermultiplet moduli $G_{ij},B_{ij}$ of the original $\mathcal N=2$ theory through  the left and right moving lattice momenta $\vec P_L, \vec P_R$. 
The shifted (2,2) Narain lattice $\Gamma_{2,2}[^H_G]$ is defined by
\begin{equation}
	\Gamma_{2,2}[^H_G](T,U) = \tau_2 \sum_{m_i,n_i\in\mathbb Z} (-1)^{mG} q^{\frac{1}{4}|P_L|^2}\,\bar q^{\frac{1}{4}|P_R|^2}\,,
	\label{HamiltonianLattice}
\end{equation}
where $P_L, P_R$ are the complex lattice momenta of the Scherk-Schwarz 2-torus parametrised in terms of the K\"ahler and complex structure moduli $T,U$ 
\begin{equation}
	P_L = \frac{m_2-Um_1+ T(n_1+Un_2)}{\sqrt{T_2 U_2}} \quad,\quad P_R = \frac{m_2-Um_1+\bar T(n_1+Un_2)}{\sqrt{T_2 U_2}} \,.
\end{equation}

Finally, the phase in the third line of \eqref{prototype} is responsible for coupling the $\mathbb Z_2'$ momentum-shift along the Scherk-Schwarz 2-torus to the fermion number parity $(-1)^{F+F_1+F_2}$ according to \eqref{ScherkSchwarzAction} and is responsible for breaking the non-abelian gauge group down to ${\rm SO}(12)\times {\rm SO}(4)\times{\rm SO}(16)$ as well as inducing the spontaneous breaking of supersymmetry. Both gravitini acquire the same mass $m_{3/2}=|U|/\sqrt{T_2 U_2}$ and we refer the reader to \cite{Angelantonj:2014dia} for more details on the model.

The $\mathcal R^2$ thresholds for this model can be organised into the following generic form
\begin{equation}
	\Delta_{\rm grav} = \frac{i}{12\pi} \int_{\mathcal F} \frac{d^2\tau}{\tau_2^2} \hat{\bar E}_2\,\frac{1}{4}\sum_{H,G}\sum_{h,g} \frac{\mathcal L[^{H,h}_{G,g}]}{\eta^2} \, \frac{\Gamma_{4,4}[^h_g]}{\eta^6\,\bar\eta^{6}}\, \Gamma_{2,2}[^H_G]\, \bar\Phi[^{H,h}_{G,g}] \,,
\end{equation}
where $\mathcal L$ is the helicity supertrace
\begin{equation}
	\mathcal L[^{H,h}_{G,g}] = \frac{1}{2}\sum_{a,b} (-1)^{a(1+G)+b(1+H)}\ \partial_\tau\left(\frac{\vartheta[^a_b]}{\eta}\right)\,\frac{\vartheta[^a_b]\,\vartheta[^{a+h}_{b+g}]\,\vartheta[^{a-h}_{b-g}]}{\eta^3} \,,
	\label{helicitysupertrace}
\end{equation}
while $\bar\Phi$ is the partition function of the gauge sector
\begin{equation}
	\bar\Phi[^{H,h}_{G,g}] = \frac{1}{\bar\eta^{18}}\, \frac{1}{2}\sum_{k,\ell=0,1} \bar\vartheta[^k_\ell]^6\,\bar\vartheta[^{k+h}_{\ell+g}]\,\bar\vartheta[^{k-h}_{\ell-g}]\, \frac{1}{2}\sum_{\rho,\sigma=0,1}\bar\vartheta[^\rho_\sigma]^8 \ (-1)^{H(\ell+\sigma)+G(k+\rho)+HG} \,.
\end{equation}
It will be useful to provide here the $T$-modular transformations of $\mathcal L$ and $\bar\Phi$
\begin{equation}
	\begin{split}
	&\mathcal L[^{H,h}_{G,g}](\tau+1) = e^{2\pi i/3 -i\pi h^2/2+i\pi H} \, \mathcal L[^{\,\,\,\,\,\,H\,\,\,\,,\,\,\,\,\,h}_{G+H\,,\,\, g+h}](\tau)\,,\\
	&\bar\Phi[^{H,h}_{G,g}](\bar\tau+1) = e^{-i\pi/2 +i\pi h^2/2+i\pi H} \, \bar\Phi[^{\,\,\,\,\,\,H\,\,\,\,,\,\,\,\,\,h}_{G+H\,,\,\, g+h}](\bar\tau)\,,
	\end{split}
	\label{modulartransf}
\end{equation}
As expected by modular covariance, the product $\mathcal L \bar\Phi$ transforms with the same constant phase regardless of the orbifold sector.

It is most convenient to organise the various contributions according to modular orbits, as outlined in \cite{Angelantonj:2013eja} and \cite{Angelantonj:2014dia, Angelantonj:2015nfa}. This allows us to consider fewer terms at the cost of enlarging the fundamental domain, which will be then invariant only under the $\Gamma_0(2)$ subgroup of ${\rm SL}(2;\mathbb Z)$, defined as
\begin{equation}
	\Gamma_0(2) = \left\{ 
			\begin{pmatrix}
				a & b\\
				c & d\\
			\end{pmatrix}\in{\rm SL}(2;\mathbb Z)\ \,{\rm and}\  c=0\,{\rm mod}\,2
	\right\} \,.
\end{equation}
 There is a trivial orbit with only one element $H=h=G=g=0$, corresponding to the original unbroken $\mathcal N=4$ theory and its contribution to the $\mathcal R^2$ term vanishes due to the appearance of additional fermionic zero modes in \eqref{helicitysupertrace}. The remaining contributions are partitioned into four ${\rm SL}(2;\mathbb Z)$ orbits
\begin{equation}
	\begin{split}
	{\rm I}\ :\ \bigr[\,^{H\,,\,h}_{G\,,\,g}\,\bigr] &= \{  \bigr[^{0\,,\,0}_{0\,,\,1}\bigr]  \} \\
	{\rm II}\ :\ \bigr[\,^{H\,,\,h}_{G\,,\,g}\,\bigr] &= \{  \bigr[^{0\,,\,0}_{1\,,\,0}\bigr]  \} \\
	{\rm III}\ :\ \bigr[\,^{H\,,\,h}_{G\,,\,g}\,\bigr] &= \{  \bigr[^{0\,,\,0}_{1\,,\,1}\bigr]  \} \\
	{\rm IV}\ :\ \bigr[\,^{H\,,\,h}_{G\,,\,g}\,\bigr] &= \{  \bigr[^{0\,,\,1}_{1\,,\,0}\bigr]+\bigr[^{0\,,\,1}_{1\,,\,1}\bigr]  \} 
	\end{split}
\end{equation}
On the r.h.s. of each line we list the $\Gamma_0(2)$ invariant elements generating each orbit under the action of the elements $S$ and $ST$ of ${\rm SL}(2;\mathbb Z)$, with $S$ being the usual inversion $\tau\to-1/\tau$ and $T$ being the unit translation $\tau\to\tau+1$. We will examine each of the above orbits separately.

\subsection{Orbit I}

Let us begin with orbit I. It contains three contributions from the sectors $H=G=0$ and $(h,g)\neq (0,0)$. Since in this orbit the Scherk-Schwarz orbifold is in its trivial sector, neither twisting nor projecting, this corresponds to a subsector of the theory with unbroken $\mathcal N=2$ supersymmetry. It essentially corresponds to half the $\mathcal N=2$ sector of the original ${\rm K}3\times T^2$ theory before the Scherk-Schwarz flux is turned on. Using the heat equation for theta functions and the Riemann identity, we can explicitly perform the sum over $(a,b)$ spin structures and evaluate  
\begin{equation}
	\mathcal L[^{0,h}_{0,g}] = -\frac{i\pi}{2}\,\eta^2\,\vartheta[^{1+h}_{1+g}]\,\vartheta[^{1-h}_{1-g}]\,.
\end{equation}
The product of theta functions in the helicity supertrace above then cancel precisely the holomorphic part of the twisted K3 lattice \eqref{K3lattice}, and so do the holomorphic Dedekind functions, so that in the end one obtains an integral of the unshifted Narain lattice times an almost (anti)holomorphic modular function
\begin{equation}
	\Delta_{\rm grav}^{\rm I}= -\frac{1}{6}\int_{\mathcal F} d\mu\ \Gamma_{2,2}(T,U)\, \hat{\bar E}_2\,\left[ \sum_{(h,g)\neq(0,0)} \frac{\bar\Phi[^{0,h}_{0,g}]}{\bar\vartheta[^{1+h}_{1+g}]\,\bar\vartheta[^{1-h}_{1-g}]}\right],
	\label{orbitI}
\end{equation}
where $d\mu=d^2\tau/\tau_2^2$ denotes the SL(2) invariant integration measure.
The quasi-holomorphy is due to the subtraction of the non-holomorphic term $3/\pi\tau_2$ from the holomorphic (but modular anomalous) Eisenstein series $E_2(\tau)$ in order to form its modular completion $\hat E_2$. It is important to mention that the quantity within brackets is fully determined by modularity, holomorphy and the knowledge of the principal part of its $q$-expansion (i.e. by its pole structure around $q=0$). It has modular weight $-2$ and is invariant under the full modular group ${\rm SL}(2;\mathbb Z)$. Modularity then implies that it can be uniquely expanded into the polynomial ring of holomorphic modular forms generated by $\{E_4,E_6,1/\Delta\}$, where $E_4, E_6$ are the holomorphic Eisenstein series of weight 4 and 6, respectively, while $\Delta=\eta^{24}$ is the modular discriminant.   The only element of the ring with weight $-2$ and a simple pole is $\bar E_4 \bar E_6/\bar\Delta=1/\bar{q}+\ldots$ and its overall coefficient may be immediately fixed by comparing the normalisation of the pole. The simple pole arising from the quantity in brackets of \eqref{orbitI} is associated to the contribution of the unphysical tachyon (protograviton) and its coefficient is found to be $-1/4$, so that
explicitly, we have
\begin{equation}
	\sum_{(h,g)\neq(0,0)} \frac{\bar\Phi[^{0,h}_{0,g}]}{\bar\vartheta[^{1+h}_{1+g}]\,\bar\vartheta[^{1-h}_{1-g}]} = -\frac{1}{4}\,\frac{\bar E_4\,\bar E_6}{\bar\Delta}\,.
\end{equation}
Therefore, the gravitational threshold contribution from orbit I is explicitly written as
\begin{equation}
	\Delta_{\rm grav}^{\rm I} = -\frac{1}{2\times 12} \int_{\mathcal F}d\mu\, \Gamma_{2,2}(T,U)\, \frac{\hat{\bar E}_2\, \bar E_4\,\bar E_6}{\bar\Delta}\,,
\end{equation}
and is recognised as precisely half the gravitational threshold of the unbroken $\mathcal N=2$ theory on ${\rm K}3\times T^2$.

\subsection{Orbit II}
We now move on to orbit II which is generated by the element $H=h=g=0$ and $G=1$. This is, in a certain sense, a universal orbit, in that it does not depend on the original $\mathcal N=2$ construction before the Scherk Schwarz breaking. Essentially, this may be seen as half the contribution of an $\mathcal N=4$ theory on $T^6$ spontaneously broken to $\mathcal N=0$ by the free $\mathbb Z_2'$ action and, therefore, it does depend on the specific way $\mathbb Z_2'$ acts on the gauge degrees of freedom. Since supersymmetry is broken in this orbit, we no longer expect to find a simplified form of the stringy Schwinger integral involving a Narain lattice times an almost holomorphic modular form. However, modularity will still allow us to organise this contribution in a way that illuminates its structure.

Orbit II contains a total of three elements, which may be obtained by acting on $[^{0,0}_{1,0}]$ by $S$ and $ST$ modular transformations. Schematically
\begin{equation}
	\Delta_{\rm grav}^{\rm II} = \int_{\mathcal F} d\mu\, (1+S+ST)\cdot \left[\begin{array}{c}
																					0 \ , \ 0\\
																					1 \ , \ 0\\
																				\end{array}\right](\tau,\bar\tau) \,.
\end{equation}
Since the measure is modular invariant, we can change variables to $\tau'=S\cdot\tau$ or $\tau'=ST\cdot\tau$ for the corresponding terms, and effectively undo these transformations at the expense, however, of enlarging the fundamental domain $\mathcal F$. Indeed, under these changes of variables $\mathcal F$ is mapped to its non-trivial images $S\cdot\mathcal F$ and $ST\cdot\mathcal F$, so that eventually, we obtain
\begin{equation}
	\Delta_{\rm grav}^{\rm II} = \int_{\mathcal F_0(2)} d\mu\, \left[\begin{array}{c}
																					0 \ , \ 0\\
																					1 \ , \ 0\\
																				\end{array}\right](\tau,\bar\tau) \,,
\end{equation}
where $\mathcal F_0(2) = \mathcal F \cup (S\cdot\mathcal F)\cup (ST\cdot \mathcal F)$. The advantage of this procedure is that we have retained only the generating element of the orbit, while the rest have been absorbed into an enlargement of the integration domain. Of course, the integrand is no longer invariant under the full modular group but only under its $\Gamma_0(2)$ subgroup. The new domain $\mathcal F_0(2) = \mathbb H^+/\Gamma_0(2)$ is exactly the fundamental domain associated to the Hecke congruence subgroup $\Gamma_0(2)$ and powerful techniques for evaluating integrals of this type were obtained in \cite{Angelantonj:2012gw,Angelantonj:2013eja,Florakis:2013ura,Angelantonj:2015rxa}. Essentially, the partial unfolding method that we have presented above expresses the obvious fact that the only necessary information needed in order to evaluate the threshold of a given orbit is already contained in its generating element. Explicitly, we have
\begin{equation}
	\Delta_{\rm grav}^{\rm II} = \frac{i}{48\pi} \int_{\mathcal F_0(2)}d\mu\, \Gamma_{2,2}[^0_1]\,\Gamma_{4,4}(G,B)\, \frac{\mathcal L[^{0,0}_{1,0}]}{\eta^8} \, \frac{\hat{\bar E}_2\, \bar\Phi[^{0,0}_{1,0}]}{\bar\eta^{6} }\,.
\end{equation}
An important difference with Orbit I, is that now the (4,4) lattice is untwisted and, therefore, there is a non-trivial contribution to the gravitational threshold involving the hypermultiplet moduli $G_{ij},B_{ij}$  of the unbroken $\mathcal N=2$ theory. In a supersymmetric setup, the $\mathcal R^2$ corrections would have been BPS saturated and the $h=g=0$ term contribution would have vanished identically due to the unsaturated fermionic zero modes in the helicity supertrace $\mathcal L$, as one would expect from the $\mathcal N=2$ structure.

With the definition \eqref{44lattice}, the (4,4) lattice carries left- and right- moving modular weight $(w,\bar w)=(2,2)$, whereas the (2,2) lattice is normalised to carry zero weight. An inspection of the transformation law \eqref{modulartransf} reveals that  $\mathcal L[^{0,0}_{1,0}]/\eta^8$ is invariant under $\tau\to\tau+1$ and is identified as a holomorphic modular form of $\Gamma_0(2)$ carrying left-moving weight $-2$. Similarly, $\bar\Phi[^{0,0}_{1,0}]/\bar\eta^6$ is invariant under $\bar\tau\to\bar\tau+1$ and therefore has to be an anti-holomorphic modular form of $\Gamma_0(2)$ with right-moving weight $-4$.

Let us note here that the holomorphic modular form $\mathcal L[^{0,0}_{1,0}]/\eta^8$ in this orbit is truly universal and does not depend in any way on the details of the construction. This is because it is essentially the helicity supertrace for Scherk-Schwarz breaking of an $\mathcal N=4$ theory and does not depend on the action of $\mathbb Z_2'$ on the gauge degrees of freedom. Any element $F$ of the ring of holomorphic modular forms of $\Gamma_0(2)$ of weight $w$ can be uniquely decomposed as
\begin{equation}
	F = A + B X_2 + C X_2^2\,,
\end{equation}
where $A,B,C$ are familiar holomorphic modular forms of ${\rm SL}(2,\mathbb Z)$ of weights $w, w-2$ and $w-4$, respectively, and $X_2(\tau) = E_2(\tau)-2E_2(2\tau)$ is the weight 2 holomorphic modular form of $\Gamma_0(2)$. Taking into account that modular forms appearing in string amplitudes may at most exhibit a simple pole at $q=0$ we may expand
\begin{equation}
	\frac{\mathcal L[^{0,0}_{1,0}]}{\eta^8} = \alpha\frac{E_4 E_6}{\Delta} + \beta \frac{E_4^2 X_2}{\Delta} + \gamma \frac{E_6 X_2^2}{\Delta} \,,
	\label{Leta}
\end{equation}
with $\alpha,\beta,\gamma$ being constants (with rational ratios) to be determined. We will now see how modularity and holomorphy will determine this modular form uniquely. As with the case of the full modular group, also for holomorphic forms of $\Gamma_0(2)$ an analogous principle holds. Namely, holomorphy and modularity uniquely fix the coefficients $\alpha,\beta,\gamma$ by the knowledge of the coefficients of the poles in the $q$-expansion, around all cusps. The fundamental domain of $\Gamma_0(2)$ contains exactly two cusps, $\tau=i\infty$ and $\tau=0$. We will, hence, study the behavior of $\mathcal L/\eta^8$ around both.

For the cusp at infinity, one may use the standard $q$-expansions for the Eisenstein and Dedekind functions appearing in \eqref{Leta}. Around $q=0$ it behaves as $\mathcal L/\eta^8 =(\alpha-\beta+\gamma)/q+\ldots$, with the dots denoting regular terms. However, the lowest possible conformal weight in the supersymmetric side of the heterotic string can at most be $-1/2$, so that the above combination $\mathcal L/\eta^8$ should never exhibit a simple pole in $q$. This translates into the condition $\beta=\alpha+\gamma$. 

The same must be true around the cusp $\tau=0$. The standard $q$-expansions of Eisenstein and Dedekind functions converge only around the cusp $\tau\to i\infty$. In order to extract the behavior of the function around $q=1$, we need to perform a $\tau$-transformation that exchanges the two cusps. The simplest such transformation is known as the Atkin-Lehner involution, or Fricke transform and it has the advantage that its action on forms of $\Gamma_0(2)$ is closed. On a modular form $f(\tau)$ of weight $w$, its action is defined as ${\rm AL}\cdot f(\tau) = (\sqrt{2}\tau)^{-w}\,f(-1/2\tau)$. Acting with it on $\mathcal L/\eta$ yields
\begin{equation}
	\begin{split}
	{\rm AL}\cdot \left[\frac{\mathcal L[^{0,0}_{1,0}]}{\eta^8} \right] &=  \frac{\alpha}{2}\frac{E_4(2\tau) E_6(2\tau)}{\Delta(2\tau)} - \frac{\beta}{4} \frac{E_4^2(2\tau) X_2(\tau)}{\Delta(2\tau)} + \frac{\gamma}{8} \frac{E_6(2\tau) X_2^2(\tau)}{\Delta(2\tau)} \\
			&= \left(\frac{\alpha}{2}+\frac{\beta}{4}+\frac{\gamma}{8}\right)\,\frac{1}{q^{2}} + 6(\beta+\gamma)\,\frac{1}{q} +\ldots
	\end{split}
	\label{ALexpans}
\end{equation}
where again the dots denote regular terms. Using now the definition of the Atkin-Lehner transform, 
\begin{equation}
	{\rm AL}\cdot \left[\frac{\mathcal L[^{H,h}_{G,g}]}{\eta^8} \right]  = \frac{(-1)^{hg}}{2} \frac{\mathcal L[^{G,g}_{H,g}](2\tau)}{\eta^8(2\tau)} \,,
	\label{ALdef}
\end{equation}
we see that it maps the original $\mathcal L/\eta^8$ in the sector $[^{0,0}_{1,0}]$ into the same function evaluated at the `dual' sector $[^{1,0}_{0,0}]$ but with doubled argument $\tau \to 2\tau$. Therefore, in order to match with the explicit expansion of $\mathcal L/\eta^8$ in the physical sector of interest, we should make sure to rescale back $\tau\to\tau/2$ so that the double pole on the r.h.s. of the second line of  \eqref{ALexpans} becomes only a simple pole in $q$, while the subleading contribution becomes a branch cut and corresponds to the familiar $-1/2$ conformal weight of the supersymmetric left-moving side of the string worldsheet. As before, we need to impose that the coefficient of the simple pole vanishes, and we obtain the second condition $4\alpha+2\beta+\gamma=0$. 

Together, these two conditions imply $(\alpha,\beta,\gamma)=(-\gamma/2,\gamma/2,\gamma)$ and we see that \eqref{Leta} has been determined up to an overall normalisation constant. It is now straightforward to fix even this overall normalisation by determining the helicity supertrace coefficient associated to the (ground state) branch cut $q^{-1/2}$ in the `dual' sector $[^{1,0}_{0,0}]$,  (the second term in the second line of \eqref{ALexpans}). Since the ground state in this sector is effectively a tachyon (unphysical), it is necessarily a scalar with multiplicity one and, therefore, its helicity supertrace contribution is $1/12$, which is to be multipled with the additional $-i\pi$ factor required by the definition of $\mathcal L$. Therefore, the coefficient of the branch cut in this sector is $-\frac{i\pi}{12}\,q^{-1/2}$ which, upon using \eqref{ALexpans} and \eqref{ALdef} yields $\gamma=-i\pi/108$, so that the full result reads
\begin{equation}
	\frac{\mathcal L[^{0,0}_{1,0}]}{\eta^8} =\frac{8i\pi}{3}\, \frac{E_4 E_6 -E_4^2 X_2 -2E_6 X_2^2}{1152\,\Delta} \,.
\end{equation}

We wish to stress that, although the above universal contribution $\mathcal L/\eta^8$ could have easily been determined in terms of Jacobi theta and Dedekind functions directly using its definition \eqref{helicitysupertrace}, the advantage of working entirely with arguments of modularity and holomorphy is clear. The modular forms arising in stringy Schwinger-like integrals in each orbit are fully determined by modularity, holomorphy, and the structure of the poles. Since the latter have specific physical meaning, there is a strong indication that the entire structure of gravitational threshold corrections is also organised into universality classes, despite the fact that the full integrand (modulo the Narain lattice) is not itself a holomorphic function of $\tau$.

Let us now use this technology in order to evaluate the right-moving contribution $\bar\Phi/\bar\eta^6$ which is model dependent, in that it depends on the action of the Scherk-Schwarz orbifold on the gauge sector. The weight now is $-4$, and we expand
\begin{equation}
	\begin{split}
	\frac{\bar \Phi[^{0,0}_{1,0}]}{\bar\eta^6} &= a \,\frac{\bar E_4^2}{\bar\Delta}+b\,\frac{\bar E_6\,\bar X_2}{\bar\Delta}+c\,\frac{\bar E_4\,\bar X_2^2}{\bar \Delta} \\
				&= (a-b+c)\,\frac{1}{\bar{q}}+(504a+456b+312c) +\ldots
	\end{split}
\end{equation}
The quantity $\bar\Phi/\bar\eta^6$ is nothing but the gauge sector contribution to the partition function, together with the right-moving oscillators coming from the $1/\bar\eta^{24}$ factor. The simple pole corresponds precisely to the unphysical tachyon of the heterotic string, and it appears with multiplicity one, hence $a-b+c=1$. As before, let us consider the Atkin-Lehner transform of this
\begin{equation}
	\begin{split}
	{\rm AL}\cdot \left[ \frac{\bar \Phi[^{0,0}_{1,0}]}{\bar\eta^6}\right] &= \frac{a}{4} \,\frac{\bar E_4^2(2\bar\tau)}{\bar\Delta(2\bar\tau)}-\frac{b}{8}\,\frac{\bar E_6(2\bar\tau)\,\bar X_2(\bar\tau)}{\bar\Delta(2\bar\tau)}+\frac{c}{16}\,\frac{\bar E_4(2\bar\tau)\,\bar X_2^2(\bar\tau)}{\bar \Delta(2\bar\tau)} \\
				&= \left(\frac{a}{4}+\frac{b}{8}+\frac{c}{16}\right)\frac{1}{\bar q^2}+3(b+c)\frac{1}{\bar q}+\ldots
	\end{split}
	\label{ALphi}
\end{equation}
The quantities of interest are the coefficients of the double and single pole, which count the multiplicities of the ground states of the right-moving sector. To compare this with a physical sector, we first note that
\begin{equation}
	{\rm AL}\cdot \left[ \frac{\bar \Phi[^{H,h}_{G,g}](\bar\tau)}{\bar\eta^6(\bar\tau)}\right] =\frac{(-1)^{hg}}{4}\,\frac{\bar\Phi[^{G,g}_{H,h}](2\bar\tau)}{\bar\eta^6(2\bar\tau)} \,.
\end{equation}
In the `dual' sector $[^{1,0}_{0,0}]$, the effective GSO projections associated to the two original ${\rm E}_8$'s are now twisted, and may produce at least massless states. In terms of SO(16) characters, the contribution is of the form $(\bar V_{16}+\bar S_{16})(\bar V_{16}+\bar S_{16})$ and therefore, neither the double nor the single pole can be present in the $\bar q$ expansion of \eqref{ALphi}, so that we obtain $4a+2b+c=0$ and $b+c=0$. Solving all three conditions, we immediately obtain $(a,b,c)=(1/9,-4/9,4/9)$ and hence,
\begin{equation}
	{\rm AL}\cdot \left[ \frac{\bar \Phi[^{0,0}_{1,0}]}{\bar\eta^6}\right] = \frac{\bar E_4^2-4\bar E_6 \bar X_2+4\bar E_4 \bar X_2^2}{9\bar \Delta} \,.
\end{equation}
Therefore, modularity and holomorphy have completely determined the Schwinger-like integrand based entirely on the structure of the poles counting the multiplicities of the ground states. With a bit more effort, we could have alternatively imposed the condition $504a+456b+312c=d_0$, where $d_0=-8$ is the multiplicity of massless right-moving states in the $[^{0,0}_{1,0}]$ sector, weighted by the appropriate parity $(-1)^{F_1+F_2}$ of the Scherk-Schwarz orbifold, and we would have obtained the exact same results.

Putting everything together, the contribution of orbit II to the gravitational thresholds reads
\begin{equation}
	\Delta_{\rm grav}^{\rm II} = -\frac{1}{18}\int_{\mathcal F_0(2)} d\mu\, \Gamma_{2,2}[^0_1]\,\Gamma_{4,4}\, \frac{E_4 E_6 -E_4^2 X_2 -2E_6 X_2^2}{1152\,\Delta} \,\frac{\hat{\bar E}_2(\bar E_4^2-4\bar E_6 \bar X_2+4\bar E_4 \bar X_2^2)}{9\bar \Delta}\,,
\end{equation} 
and we have separately fixed the normalisation of the holomorphic and anti-holomorphic parts inside the integral to unity.

\subsection{Orbit III}

We move on to examine orbit III, defined by $(h,g)=(H,G)\neq(0,0)$. Effectively, this sits at the overlap between the two $\mathbb Z_2$ orbifold factors and is actually supersymmetric due to the de-corellation of the R-symmetry charge to the (2,2) lattice. This can be seen immediately from the helicity supertrace formula
\begin{equation}
	\mathcal L[^{H,H}_{G,G}] = \frac{1}{2}\sum_{a,b} (-1)^{a(1+G)+b(1+H)}\ \partial_\tau\left(\frac{\vartheta[^a_b]}{\eta}\right)\,\frac{\vartheta[^a_b]\,\vartheta[^{a+H}_{b+G}]\,\vartheta[^{a-H}_{b-G}]}{\eta^3} \,,
\end{equation}
upon recognising that the phase can be split into the spin statistics factor $(-1)^{a+b}$ and the additional factor $(-1)^{(a+H)(b+G)+ab+HG}$. Aside from the irrelevant for our argument $(-1)^{HG}$, the additional phase factor only contributes whenever $a+H=b+G=1$. But in this case, the entire contribution vanishes due to the remaining theta factors in the sum. The extra phase may then be replaced by $(-1)^{HG}$ and the R-symmetry charges $(a,b)$ decouple entirely from the Scherk-Schwarz lattice. This subsector is then interpreted as (one half) the original $\mathcal N=4$ theory spontaneously broken to $\mathcal N=2$ by the single orbifold action
\begin{equation}
	v = (-1)^{F_1+F_2} \,r\,\delta\,,
\end{equation}
where $r$ is the crystallographic rotation $Z^{1,2}\to -Z^{1,2}$ acting on the complexified $T^4$ super-coordinates and $\delta$ again realises the momentum shift along $Z^3$. As with all supersymmetric cases, the holomorphic dependence cancels out between the helicity supertrace and the twisted (4,4) lattice
\begin{equation}
	\mathcal L[^{H,H}_{G,G}] = \frac{i\pi (-1)^{HG+1}}{2}\, \eta^2\, \vartheta[^{1+H}_{1+G}]\,\vartheta[^{1-H}_{1-G}]\,.
\end{equation}

 With partial unfolding from ${\rm SL}(2;\mathbb Z)$ to its $\Gamma_0(2)$ subgroup, as in the previous subsection, we may cast the contribution to the gravitational thresholds from orbit III into the form
\begin{equation}
	\Delta_{\rm grav}^{\rm III} = -\frac{1}{6} \int_{\mathcal F_0(2)}d\mu\, \Gamma_{2,2}[^0_1]\, \hat{\bar E}_2\, \frac{\bar\Phi[^{0,0}_{1,1}]}{\bar\vartheta_2^2} \,.
\end{equation}
Working in a similar way as for orbit II, we identify the contribution $\bar\Phi/\bar\theta_2^2$ as an anti-holomorphic modular form of weight $-2$ and we can again fix its expansion in terms of the multiplicities of the poles. The result for the contribution of orbit III is
\begin{equation}
	\Delta_{\rm grav}^{\rm III} = -\frac{1}{2\times 12}\int_{\mathcal F_0(2)} d\mu\, \Gamma_{2,2}[^0_1]\ \frac{\hat{\bar E}_2 (\bar E_4 \bar E_6-4 \bar E_4^2 \bar X_2+4 \bar E_6 \bar X_2^2)}{9\bar \Delta} \,.
\end{equation}

\subsection{Orbit IV}

As with orbit II, also orbit IV generated by the elements $H=0$, $h=G=1$ summed over $g=0,1$ is manifestly non-supersymmetric and, therefore, its holomorphy properties may a priori seem obscured in expressions that at first sight appear quite involved. They may, however, be disentangled using partial unfolding and organised again according to the principles of modularity and holomorphy using the knowledge of the multiplicities of ground states (unphysical tachyons) in various sectors. To see this, we begin by unfolding the integral on the ${\rm SL}(2;\mathbb Z)/\Gamma_0(2)$ coset, and find
\begin{equation}
	\Delta_{\rm grav}^{\rm IV} = \frac{i}{3\pi} \int_{\mathcal F_0(2)}d\mu \ \Gamma_{2,2}[^0_1]\, \hat{\bar E}_2\, \left[\sum_{g=0,1}\frac{\mathcal L[^{0,1}_{1,g}]}{\eta^2\,\vartheta[^{\ \,0\,}_{1+g}]^2} \times \frac{\bar\Phi[^{0,1}_{1,g}]}{\bar\vartheta[^{\ \,0\,}_{1+g}]^2} \right]\,.
	\label{orbIV}
\end{equation}
Naturally, each term separately in the sum does not have well-defined modular properties, even though the sum does. The quantity in brackets in \eqref{orbIV} admits the following diagonal decomposition
\begin{equation}
\frac{1}{2}\left[\sum_{g=0,1}\frac{\mathcal L[^{0,1}_{1,g}]}{\eta^2\,\vartheta[^{\ \,0\,}_{1+g}]^2}\right]\left[\sum_{g'=0,1}\frac{\bar\Phi[^{0,1}_{1,g'}]}{\bar\vartheta[^{\ \,0\,}_{1+g'}]^2}\right] + \frac{1}{2}\left[\sum_{g=0,1}(-1)^g\frac{\mathcal L[^{0,1}_{1,g}]}{\eta^2\,\vartheta[^{\ \,0\,}_{1+g}]^2}\right] \times \left[\sum_{g'=0,1}(-1)^{g'}\frac{\bar\Phi[^{0,1}_{1,g'}]}{\bar\vartheta[^{\ \,0\,}_{1+g'}]^2}\right] \,,
\label{decompOrbIV}
\end{equation}
and simplifies even further by noticing that
\begin{equation}
	\sum_{g=0,1}\frac{\bar\Phi[^{0,1}_{1,g}]}{\bar\vartheta[^{\ \,0\,}_{1+g}]^2} =-\frac{1}{\bar\eta^{12}}\sum_{g=0,1}\sum_{\rho,\sigma=0,1}\left[\sum_{k,\ell=0,1}(-1)^{k+\ell+\rho}\frac{\bar\vartheta[^k_\ell]^4}{\bar\vartheta[^{\,\,\,\,0\,}_{1+g}]^4}-(-1)^{\rho+g}\right]=0 \,,
\end{equation}
where the first term vanishes\footnote{This is a property of the particular choice of Scherk-Schwarz action on the gauge degrees of freedom in this model and not true in general. In models where $\mathbb Z_2'$ acts as $v'=(-1)^{F+F_2}\delta$, this term is in general non-vanishing.} due to a hidden spectral flow in the bosonic side of the heterotic theory (Jacobi identity), while the second term vanishes from the summation over $g$. Therefore, we can finally obtain the desired holomorphic factorisation of the integrand
\begin{equation}
	\Delta_{\rm grav}^{\rm IV} = \frac{i}{3\pi} \int_{\mathcal F_0(2)}d\mu \ \Gamma_{2,2}[^0_1]\, \hat{\bar E}_2\, \left[\sum_{g=0,1}(-1)^g\frac{\mathcal L[^{0,1}_{1,g}]}{\eta^2\,\vartheta[^{\ \,0\,}_{1+g}]^2}\right] \times \left[\frac{\bar\Phi[^{0,1}_{1,0}]}{\bar\vartheta_2^2} \right]\,.
\end{equation}
Now both the holomorphic as well as the anti-holomorphic parts within the brackets are separately odd under $\tau\to\tau+1$ and, hence, they cannot be invariant under $\Gamma_0(2)$. However, consider multiplying (and dividing) by $\eta^{12}$ which is itself odd under $\tau\to\tau+1$, and similarly with $\bar\eta^{12}$
\begin{equation}
	\Delta_{\rm grav}^{\rm IV} = \frac{i}{3\pi} \int_{\mathcal F_0(2)}d\mu \ \Gamma_{2,2}[^0_1]\, \hat{\bar E}_2\,\, \left[\sum_{g=0,1}(-1)^g\frac{\mathcal L[^{0,1}_{1,g}]}{\eta^{14}\,\vartheta[^{\ \,0\,}_{1+g}]^2}\right] \times \left[\frac{\bar\Phi[^{0,1}_{1,0}]}{\bar\eta^{12}\,\bar\vartheta_2^2} \right]\times (\eta^{12}\,\bar\eta^{12})\,.
\end{equation}
This manipulation has the advantage of turning the quantities within the brackets into modular forms of $\Gamma_0(2)$ and, therefore, they can again be uniquely fixed using holomorphy, modularity and the physical input from the low lying spectrum, such as  multiplicities of poles (or cuts) in $q$ and $\bar q$, or using information from the massless spectrum corresponding to constant terms in the expansions.

We shall not repeat here the analysis in detail, which is carried out in precisely the same way as we outlined for orbit II, but  simply present the result for the final orbit IV
\begin{equation}
	\Delta_{\rm grav}^{\rm IV} = \frac{1}{9\times 12} \int_{\mathcal F_0(2)}d\mu \ \Gamma_{2,2}[^0_1]\, \frac{E_6+E_4 X_2}{\eta^{12}}\,\, \frac{\hat{\bar E}_2(4\bar X_2^2-\bar E_4)}{3\bar\eta^{12}}\,.
\end{equation}

Although computed in a specific prototype model, the moduli-dependent structure of the threshold corrections when analysed in terms of modular orbits is generic. Had we realised K3 as another $T^4/\mathbb Z_N$ orbifold with $N=3,4,6$ we would still have obtained similar orbit decompositions with appropriate holomorphic factorisation in the corresponding integrands, involving again modular forms of $\Gamma_0(2)$ or ${\rm SL}(2,\mathbb Z)$ that are uniquely determined from the ground states of the theory by matching the structure of the poles as we did above. In particular, the above analysis continues to hold in more realistic heterotic models where $\mathcal N=1$ supersymmetry is spontaneously broken \`a la Scherk-Schwarz, such as the chiral models of \cite{Florakis:2016ani}. 

In such cases, there typically exist several different copies of the above orbits. In particular, orbit I is special in that it corresponds to the contribution of the original $\mathcal N=2$ supersymmetric theory. It is the only contribution where the (2,2) lattice associated to the Scherk-Schwarz $T^2$ is both unshifted and unprojected, $\Gamma_{2,2}[^0_0](T,U)\equiv\Gamma_{2,2}(T,U)$ and we will see it plays a central role to the behaviour of gravitational thresholds at large volume.

\section{Universality and Large volume behaviour}\label{SecLargeVol}

In order to further analyse the behaviour of the gravitational threshold $\Delta_{\rm grav}$ we must first decide in which region of the $(T,U)$ moduli space we are interested in. This is necessary because different techniques of evaluation may produce fast or slowly converging expressions for the result, depending on the region of moduli space one is working in. For instance, the method of lattice unfolding \cite{McClain:1986id,O'Brien:1987pn,Dixon:1990pc} is best applied when one is interested in the large volume $T_2\gg 1$ expansion of Schwinger-like integrals, while collapsing due to absolute convergence issues around self-dual points. On the other hand the more recent techniques in \cite{Angelantonj:2011br,Angelantonj:2012gw,Angelantonj:2013eja,Florakis:2013ura,Angelantonj:2015rxa,Florakis:2016boz} may also produce fast converging expressions valid around self-dual points.

The specific model used as a paradigm for our analysis has a negative one-loop contribution to its effective potential, and the dynamics lead the theory away from the large volume regime. This is not a phenomenon particular to the $\mathcal N=2$ theories. The structure of the one-loop effective potential around the string scale is highly model-dependent and crucially depends on the contribution of non-level matched states which have no field theory analogue. However, a wide class of chiral heterotic models with spontaneously broken $\mathcal N=1$ supersymmetry was recently analysed in \cite{Florakis:2016ani}, and the first explicit examples of stringy constructions with spontaneous decompactification to the large volume regime were presented. 

Motivated by this result, we wish to extract the general behaviour of gravitational thresholds in such models in the limit of large volume $T_2\gg 1$. Note that, even though the dynamics of our prototype model does not drive the theory to the large volume regime, nevertheless, the analysis of the realistic models of \cite{Florakis:2016ani} follows in an almost identical way. Moduli dependent corrections may only arise from $\mathcal N=2$ sectors, which take precisely the form of our simple model\footnote{Or several copies thereof, arising as the various different $\mathcal N=2$ subsectors in the $\mathcal N=1\to 0$ models of \cite{Florakis:2016ani}.}, with a slightly different choice of Scherk-Schwarz action on the gauge degrees of freedom. This implies that all the basic ingredients of our analysis, such as the decomposition of contributions into modular orbits, the principle of holomorphic factorisation  and the determination of the corresponding modular forms of $\Gamma_0(2)$ or ${\rm SL}(2,\mathbb Z)$ from the multiplicities of the ground states of the theory, may be carried over to the realistic cases with minor modifications that enter into a set of constants determining the multiplicities of the ground states of the model.

With the understanding that universality essentially expresses modularity and holomorphic factorisation, it is possible to work out the moduli-dependent part of $\mathcal R^2$ threshold corrections at one loop for this class of models and show that they may be cast into the following \emph{universal} form
\begin{equation}
\Delta_{\rm grav}=\sum_{i=1,2,3}\left(\Delta_{{\rm grav},i}^{\rm I}+\Delta_{{\rm grav},i}^{\rm II}+\Delta_{{\rm grav},i}^{\rm III}+\Delta_{{\rm grav},i}^{\rm IV}\right)\,,
\end{equation}
where
\begin{equation}
	\begin{split}
	\Delta_{{\rm grav},i}^{\rm I} &=  -\frac{\zeta_i}{12}\int_{\mathcal F}d\mu\, \Gamma_{2,2}^{(i)}(T^{(i)},U^{(i)})\, \frac{\hat{\bar E}_2\, \bar E_4\,\bar E_6}{\bar\Delta}\,,\\
	\Delta_{{\rm grav},i}^{\rm II} &= \int_{\mathcal F_0(2)} d\mu\, \Gamma_{2,2}^{(i)}[^0_1]\,\Gamma_{4,4}^{(\neq i)}\, \frac{E_4 E_6 -E_4^2 X_2 -2E_6 X_2^2}{\Delta} \,\frac{\hat{\bar E}_2(\alpha_i\,\bar E_4^2+\beta_i\,\bar E_6 \bar X_2+\gamma_i\,\bar E_4 \bar X_2^2)}{\bar \Delta}\,,\\
	\Delta_{{\rm grav},i}^{\rm III} &= \int_{\mathcal F_0(2)} d\mu\, \Gamma_{2,2}^{(i)}[^0_1]\ \frac{\hat{\bar E}_2 (\alpha_i'\,\bar E_4 \bar E_6+\beta_i'\, \bar E_4^2 \bar X_2+\gamma_i'\, \bar E_6 \bar X_2^2)}{\bar \Delta} \,,\\
	\Delta_{{\rm grav},i}^{\rm IV} &=  \int_{\mathcal F_0(2)}d\mu \ \Gamma_{2,2}^{(i)}[^0_1]\, \left(\frac{E_6+E_4 X_2}{\eta^{12}}\,\, \frac{\hat{\bar E}_2(\alpha_i''\,\bar E_4+\beta_i''\bar X_2^2)}{\bar\eta^{12}} \right. \\
	&\qquad\qquad\qquad\quad+\left.(j_0-8)\,\frac{\hat{\bar E}_2(\alpha_i''' \,\bar E_4\bar E_6+\beta_i''' \,\bar E_4^2 \bar X_2+\gamma_i''' \,\bar E_6 \bar X_2^2)}{\bar \Delta}\right)\,,
	\end{split}
	\label{GravUniversality}
\end{equation}
where the model dependence enters only into the constant coefficients $\zeta,\alpha,\beta,\gamma$ (and their primes) which count the multiplicities of the ground states of the model and may be uniquely determined as we described in the previous section. Here, we denote by $j_0(\tau)=24+(\vartheta_2(\tau)/\eta(\tau))^{12}$ the Atkin-Lehner transform of the Hauptmodul $j_2(\tau)$ of $\Gamma_0(2)$ and the index $i=1,2,3$ runs over the three 2-tori of the internal space.

Given the universal expression \eqref{GravUniversality} we may now begin to investigate the structure of gravitational $\mathcal R^2$ corrections for the class of models in \cite{Florakis:2016ani}. Following the latter, we  assume all moduli to be fixed at their fermionic point values, except for the Scherk-Schwarz 2-torus and consider the large volume limit $T_2\gg 1$. With this assumption, the only case of interest yielding $T_2$-dependent contributions to the gravitational thresholds is of the form \eqref{GravUniversality} with the (2,2) lattice being always associated to the Scherk-Schwarz 2-torus, and the (4,4) lattice in $\Delta_{\rm grav}^{\rm II}$ being expressed entirely in terms of left- and right- moving level-1 theta characters. 

We shall begin by first treating contributions II, III and IV which always involve shifted Narain lattices. They are all of the generic form
\begin{equation}
	I_{\rm} = \int_{\mathcal F_0(2)}d\mu\ \Gamma_{2,2}[^0_1]\, f[^0_1](\tau,\bar\tau)\,,
	\label{shiftIntegral}
\end{equation}
where the $\Gamma_0(2)$ modular form $f[^0_1]$ has the generic expansion
\begin{equation}
	f[^0_1](\tau,\bar\tau) = \sum_{\ell=0,1}\sum_{N,M} c_\ell(N,M)\,\tau_2^{-\ell}\, q^{N} \bar q^{M}\,.
	\label{Fourier}
\end{equation}
Although \eqref{shiftIntegral} may be explicitly computed as an asymptotic expansion in $T_2$, we shall instead focus on extracting its large volume behaviour, by keeping only the dominant contributions. This is done most conveniently by writing the shifted (2,2) lattice in its Lagrangian representation 
\begin{equation}
	\Gamma_{2,2}[^0_1] = T_2\sum_{m_i,n_i} e^{-2\pi i T\,\det A-\frac{\pi T_2}{\tau_2 U_2}\left|(1,U)A\binom{\tau}{1}\right|^2} \,,
	\label{LagrangianLat}
\end{equation}
where $A$ is the matrix of windings
\begin{equation}
	A= \begin{pmatrix}
					n_1 & m_1+\frac{1}{2}\\
					n_2 & m_2\\
	\end{pmatrix}\,.
\end{equation}
It is obtained after Poisson resumming the Kaluza-Klein momenta $m_1,m_2\in\mathbb Z$ in \eqref{HamiltonianLattice}.
Due to the lattice shift, the matrix $A$ is always non-vanishing and there is no constant volume term in the Lagrangian lattice sum. The full result may be suitably decomposed into modular orbits and the fundamental domain may be unfolded to the strip $\mathcal S$ (when $\det A=0$) or the entire upper half-plane $\mathbb H^+$ (when $\det A\neq 0$). The non-degenerate case, for which $(n_1,n_2)\neq(0,0)$, is always exponentially suppressed in $T_2$ and will be dropped for the purposes of our analysis. We are therefore left with the degenerate orbit $n_1=n_2=0$ to be integrated over the half-infinite strip $\mathcal S=\{\tau\in \mathbb{H}^+\,,\,|\tau_1|\leq\frac{1}{2}\}$. Performing explicitly the $\tau_1$ integral imposes level-matching and retains only modes with $N=M$ in the Fourier expansion \eqref{Fourier} so that we are left with
\begin{equation}
	I \simeq T_2\sum_{N,\ell}\sum_{m_1,m_2} c_\ell(N,N) \int_0^\infty \frac{dt}{t^{1+s+\ell}}\,e^{-\frac{\pi T_2}{t U_2}\left|m+\frac{1}{2}+Um_2\right|^2-4\pi N t} \,,
	\label{Schwinger}
\end{equation}
where we now denote the Schwinger parameter as $t$, and introduce the complex parameter $s$ in order to deal with IR divergences arising from the massless states ($N=0$). Indeed, the integral converges for ${\rm Re}(s)>1$, while exhibiting a simple pole at $s=1$. Since we will not deal with moduli independent constants which, at any rate, also depend on the choice of renormalisation scheme, we will simply define the integral by evaluating it for ${\rm Re}(s)>1$ and then analytically continue to $s=1$ after properly subtracting the pole. We refer the reader to \cite{Angelantonj:2011br,Angelantonj:2012gw,Angelantonj:2013eja,Florakis:2013ura,Angelantonj:2015rxa,Florakis:2016boz} for further details.

For $N\neq 0$ one recognises from \eqref{Schwinger} the integral representation of the modified Bessel function of the second kind $K_{s+\ell}(4\pi\sqrt{N T_2/U_2}|m_1+\frac{1}{2}+Um_2|)$, dressed up with additional polynomial factors. For large $T_2\gg 1$ this drops exponentially fast and shall be therefore neglected. As expected, the leading contribution to the integral in the large volume limit\footnote{This should be contrasted with the behaviour of stringy Schwinger integrals around self-dual points, where the dominant contribution to amplitudes often comes from unphysical states (see for example \cite{Florakis:2016ani}).} comes from the massless modes $N=0$. 

Evaluating the Schwinger-like integral \eqref{Schwinger} for $N=0$ and carefully extracting the pole in $s$ arising from the  constant term $\ell=0$, we find
\begin{equation}
	I = -c_0(0,0)\, \log T_2 U_2 \,|\vartheta_2(U)|^4 + \frac{2\,c_1(0,0)}{\pi^2\,T_2} \left[ 4E(2,2U)-E(2,U)\right] +\ldots \,,
	\label{SchwingerResult}
\end{equation}
where the dots stand for exponentially suppressed terms and $E(s,\tau)$ is the weight 0, real analytic Eisenstein series of ${\rm SL}(2;\mathbb Z)$
\begin{equation}
	E(s,\tau) = \frac{1}{2}\sum_{(m,n)\neq(0,0)} \frac{\tau_2^s}{|m-\tau n|^{2s}}\,.
\end{equation}
In what follows, we shall be drop also the polynomially suppressed term and focus only on the leading behaviour. 

Let us now comment on the integral of case I, which differs from the rest in that it involves the unshifted Narain lattice. Its Lagrangian representation is again of the form \eqref{LagrangianLat}, except for the fact that the winding matrix $A$ now has unshifted integer entries
\begin{equation}
	A= \begin{pmatrix}
					n_1 & m_1\\
					n_2 & m_2\\
	\end{pmatrix}\,.
\end{equation}
In particular, this implies that there is now a non-trivial contribution from the orbit $A=0$, still to be integrated over the fundamental domain $\mathcal F$ of ${\rm SL}(2,\mathbb Z)$. Importantly, this contribution is linear in the volume $T_2$ of the Scherk-Schwarz 2-torus. The next contribution comes, as before, from the degenerate orbit $n_1=n_2=0$, which is integrated over the strip in a similar way, and requires properly deforming the integral using the complex parameter $s$  and carefully extracting the finite part. Finally, there is the exponentially suppressed contribution of the non-degenerate orbit which we shall neglect. Therefore, we find
\begin{equation}
	\int_{\mathcal F}d\mu\,\Gamma_{2,2}(T,U)\, f(\tau,\bar\tau) \simeq \frac{\pi c_0(0,0)}{3}T_2-c_0(0,0)\,\log T_2 U_2\,|\eta(U)|^4+\ldots \,,
\end{equation}
where we again assumed the generic form \eqref{Fourier} for the Fourier expansion of the ${\rm SL}(2,\mathbb Z)$ modular form $f(\tau,\bar\tau)$.

Putting everything together, we can now give the leading behaviour of the gravitational thresholds \eqref{GravUniversality} in each of the four cases as
\begin{equation}
	\begin{split}
				&\Delta^{\rm I} = \frac{22\pi\zeta}{3}\,T_2 - 22\zeta\log T_2 +\ldots \,,\\
				&\Delta^{\rm II}= -110592 (5\alpha+5\beta+3\gamma)\, \log T_2 +\ldots \,,\\
				&\Delta^{\rm III}= 24(11\alpha'+21\beta'+19\gamma')\,\log T_2 + \ldots \,,\\
				&\Delta^{\rm IV} = 384(11\alpha'''+21\beta'''+19\gamma''')\,\log T_2+\ldots \,, 
	\end{split}
	\label{ThreshLargeVol}
\end{equation}
where the dots again denote sub-leading terms. An alternative way to interpret the coefficients in \eqref{ThreshLargeVol} is by identifying them as trace anomaly coefficients for the massless spectrum of the theory, matching with \cite{Antoniadis:1992sa,Antoniadis:1992rq,Gates:1983nr}.

 The contributions in II, III and IV reproduce the logarithmic growth in $T_2$ ascribed to states with masses below the Kaluza-Klein scale, as expected from field theory. More interesting is the contribution of $\Delta^{\rm I}$. It shows that at large volume, the gravitational thresholds are dominated by the linear growth in $T_2$ arising from the first term in $\Delta^{\rm I}$. In the decompactification limit, the physics becomes effectively six-dimensional and the first line of \eqref{ThreshLargeVol} is essentially removing the logarithmic growth and replacing it with a linear growth in the volume, in accordance with scaling arguments.

Before closing, we wish to emphasise a point which becomes especially important in models with spontaneously broken supersymmetry. The presence of the linear growth in $\Delta^{\rm I}$ is ubiquitous in heterotic orbifolds with $\mathcal N=2$ sectors having fixed points. Such is the case, for example, for the prototype model of section \ref{prototypeSection}. Namely, for a generic theory with spontaneous supersymmetry breaking \`a la Scherk-Schwarz, we may always focus on the subsector $H=G=0$ where the breaking is absent. If the resulting theory is an $\mathcal N=4$ one, then of course no contribution arises to the $\mathcal R^2$ thresholds from this sector. Another possibility is that the theory in this subsector enjoys $\mathcal N=2$ supersymmetry, but is itself obtained by partial spontaneous breaking from an $\mathcal N=4$ one. In this case, the running of the gravitational couplings is logarithmic, as in $\Delta^{\rm III}$, due to the presence of enhanced $\mathcal N=4$ supersymmetry in the infinite volume limit.

 If, however, the theory in the $H=G=0$ sector contains `true' $\mathcal N=2$ sub-sectors with fixed points (i.e. there is no enhancement to $\mathcal N=4$ as $T_2\to\infty$), then the moduli dependent contributions to $\mathcal R^2$, will contain a contribution $\Delta^{\rm I}$ involving the unshifted Narain lattice associated to the Scherk-Schwarz 2-torus, and will generically grow linearly with the volume, as depicted in the first line of \eqref{ThreshLargeVol}. As a result, the gravitational sector will be quickly driven to strong coupling. This is the gravitational analogue of the decompactifaction problem for gauge couplings \cite{Kiritsis:1996xd}, which is particularly important in models with spontaneous supersymmetry breaking, since the tree-level no-scale moduli typically acquire a potential at one-loop. Chiral models which dynamically provide an attractor \cite{Florakis:2016ani} to low supersymmetry breaking scales via Scherk-Schwarz ($T_2\gg 1$), will generically induce such a linear $T_2$-behaviour in the running of gauge as well as gravitational couplings. 
 
It is hoped that further analysis of threshold corrections as well as corrections to the effective potential, supplemented by an appropriate stabilisation mechanism as in \cite{Abel:2016hgy} might offer a natural resolution.



\section*{Acknowledgements}

We wish to thank C.~Angelantonj, C.~Condeescu, J.~Rizos, J.~Troost and especially C.~Kounnas for useful discussions. We would like to further acknowledge the University of Athens for its warm hospitality during the final stages of this work.


\bibliographystyle{utphys}
\providecommand{\href}[2]{#2}\begingroup\raggedright\endgroup

\end{document}